\newif\ifloguseIEEEConf
\ifloguseIEEEConf
    \documentclass[10pt,conference]{IEEEtran}
\else
    \documentclass{article}
\fi

\usepackage{graphicx}
\usepackage{lscape}
\usepackage{amssymb}
\usepackage[letterpaper,top=3cm,bottom=3.25cm,left=2.75cm, right=2.75cm]{geometry}
\usepackage{amsfonts,amsmath, amssymb, graphics,geometry,framed,enumitem}

\begin{document}

\newtheorem{theorem}{Theorem}
\newtheorem{corollary}{Corollary}
\newtheorem{lemma}{Lemma}
\newtheorem{example}{Example}
\newtheorem{definition}{Definition}
\newtheorem{proposition}{Proposition}
\newtheorem{observation}{Observation}
\newtheorem{conjecture}{Conjecture}
\newtheorem{remark}{Remark}
\newcommand{\N}{\mathbb{N}}
\newcommand{\qed}{\hfill $\diamondsuit$}
\newcommand{\noam}{Noam Presman\ }
\newcommand{\bhat}{Bhattacharyya }
\newcommand{ \etld } {\tilde{\epsilon}}
\newcommand{ \He } {\mathcal{H}}

\ifloguseIEEEConf

\else
  \newcommand{\QED}{\hfill $\diamondsuit$}
  \newcommand{\proof}{\noindent {\bf Proof}\ }
\fi

\title{Binary Polar Code Kernels from Code Decompositions  }

\author{Noam Presman, Ofer Shapira and Simon Litsyn\\ School of Electrical Engineering,
Tel Aviv University, Ramat Aviv 69978 Israel. \\e-mails:
 \{presmann, ofershap, litsyn\}@eng.tau.ac.il. }
\date{}

\maketitle

\begin{abstract} Code decompositions (a.k.a code nestings) are used to
design good binary polar code kernels. The proposed kernels are in
general non-linear and show a better rate of polarization under
\textit{successive cancelation} decoding, than the ones suggested by
Korada et al., for the same kernel dimensions. In particular,  kernels of sizes 14, 15 and 16 are constructed and shown to provide  polarization
rates better than any binary kernel of such sizes.
\end{abstract}
\section{Introduction}
Polar codes were introduced by Arikan \cite{Arikan} and provided
a scheme for achieving the symmetric capacity of binary
memoryless channels (B-MC) with polynomial encoding and decoding
complexity. Arikan used  a simple construction based on the
following linear kernel
$$
G_2 = \left(
      \begin{array}{cc}
        1 & 0 \\
        1 & 1 \\
      \end{array}
    \right).
$$
In this scheme,  a $2^n\times2^n$ matrix, $G_2^{\bigotimes n}$, is generated by performing the  Kronecker power on $G_2$. An input vector $\bf u$ of length   $N=2^n$  is transformed to an $N$ length vector $\bf x$ by multiplying a certain permutation of the vector $\bf u$ by $G_2^{\bigotimes n}$. The vector $\bf x$ is  transmitted  through  $N$ independent copies of the memoryless channel, $W$.  This results in new $N$ (dependent) channels between the individual components of $\bf u$  and the outputs of the channels. Arikan showed that these
channels exhibit the phenomenon of polarization under successive
cancelation decoding. This means that as $n$ grows there is a
proportion of $I(W)$ (the symmetric channel capacity) of the
channels that become clean channels (i.e. having the capacity
approaching $1$) and the rest of the channels become completely
noisy (i.e. with the capacity approaching $0$). An important
question is how fast the polarization occurs in terms of the codes'
length $N$. In \cite{Arikan2}, the rate of polarization was analyzed
for the $2\times 2$ kernel, and it was proven that the rate is
$O\left(2^{-N^{0.5}}\right)$. More specifically the authors showed that
\begin{equation}\label{eq:introRate1}
\liminf_{n\rightarrow \infty}\Pr\left(Z_n\leq 2^{-N^\beta}\right)=I(W)
\,\,\,\, \text{for}\,\,\, \beta<0.5
\end{equation}
\begin{equation}\label{eq:introRate2}
\liminf_{n\rightarrow \infty}\Pr\left(Z_n\geq 2^{-N^\beta}\right)=1
\,\,\,\,\text{for}\,\,\,\, \beta>0.5, \end{equation}
where $\left\{Z_n\right\}_{n\geq 0}$ is the Bhattacharyya random sequence corresponding to Arikan's random tree process \cite{Arikan}.

In \cite{Korada}, Korada \textit{et al.} studied the use of alternatives
to $G_2$ for the symmetric B-MC. They gave sufficient conditions for
polarization when linear binary kernels are used over the symmetric
B-MC channels. Furthermore,    the notion of the rate of
polarization was generalized  for polar codes based on linear codes
having generating matrix $G$ of dimensions ${\ell} \times {\ell}$.
The rate of polarization was quantified by the  exponent of the
kernel $E(G)$, which plays the general role of the threshold (equal
$0.5$) appearing in (\ref{eq:introRate1}) and
(\ref{eq:introRate2}) (note that here $N={\ell}^n$).  Korada \textit{et al.}
showed that $E(G)\leq 0.5$ for all binary linear kernels of
dimension ${\ell}\leq 15$, which is the kernel exponent found for
Arikan's $2\times 2$ kernel, and that for ${\ell}=16$ there exists a
code generator matrix $G$ in which $E(G)= 0.51828$, and this is the
maximum exponent achievable by a binary linear kernel up to this
dimension. Furthermore, for optimal linear
kernels, the exponent $E(G)$ approaches 1 as ${\ell}\rightarrow
\infty$.

In \cite{MoriandTanka}, Mori and Tanaka considered the general case
of a mapping $g(\cdot)$, which is not necessarily linear and binary,
as a basis for channel polarization constructions. They gave
sufficient conditions for polarization and generalized the exponent
for these cases. In \cite{MoriandTanka3}, they considered non-binary,
however linear, kernels based on Reed-Solomon codes and Algebraic
Geometry codes and showed that their exponents are by far better
than the exponents of the known binary kernels. This is true even
for such a small kernel dimension as ${\ell}=4$ and the alphabet
size $q=4$, in which $E\left(G\right)=0.573120$.

In this paper, we propose designing  good binary kernels
(in the sense of large exponent), by using code decompositions
(a.k.a code nestings). The kernels we suggest show better exponents
than the ones considered in \cite{Korada}. Moreover, we describe
binary non-linear kernels of sizes 14, 15 and 16 providing a superior
polarization exponent than any binary linear kernel.

The paper is organized as follows. In Section
\ref{sec:CodesDecom}, we describe building kernels  that are related
to decompositions of codes into sub-codes. Furthermore, by using results from \cite{MoriandTanka}, we observe that the exponent of these kernels is a function of the partial minimum distances between the sub-codes.  We then develop in Section \ref{sec:Bounds} an upper-bound on the exponent of dimension $\ell$. In Section
\ref{sec:ExmpOfGoodKernels}, we give examples of known code decompositions which result in binary kernels that achieve the upper-bounds from Section \ref{sec:Bounds}.

\section{Preliminaries}\label{sec:CodesDecom}
We consider  kernels that are based on bijective binary
transformations. A channel polarization kernel of dimension
${\ell}$, denoted by $g(\cdot)$, is a bijective mapping
$$
g:\left\{0,1\right\}^{{\ell}}\rightarrow \left\{0,1\right\}^{{\ell}}.
$$
This means that $g({\bf u})={\bf x}, \,\,\,\, {\bf u}, {\bf x}
\in\left\{0,1\right\}^{{\ell}}$. Denote the output components of the
transformation by
$$
g_i({\bf u})=x_i      \,\,\,\ i\in[{\ell}],
$$
where for a natural number ${\ell}$, we denote
$[{\ell}]=\left\{1,2,3,...,{\ell}\right\}$. For $i\geq j$, let ${\bf
u}^i_j=(u_j,...,u_i)$ be the sub-vector of ${\bf u}$ of length
$i-j+1$ (if $i<j$ we say that ${\bf
u}^i_j=()$, the empty vector, and its length is $0$). It is convenient to denote by $g^{({\bf
v}_1^i)}:\left\{0,1\right\}^{{\ell}-i}\rightarrow
\left\{0,1\right\}^{{\ell}}$, the restriction of $g(\cdot)$ to the
set $\left\{{\bf v}_1^i {\bf u}_1^{{\ell}-i}| {\bf
u}_1^{{\ell}-i}\in \left\{0,1 \right\}^{{\ell}-i}\right\}$, that is
$$
g^{({\bf v}_1^i)}({\bf u}_1^{{\ell}-i}) =g({\bf v}_1^i {\bf
u}_1^{{\ell}-i}) \,\,\,\,\, i \in [{\ell}-1].
$$
 Next, we consider code decompositions. The initial code is
partitioned to several sub-codes having the same size. Each of these
sub-codes can be further partitioned. Here we choose as the initial
code, the total space of  length ${\ell}$ binary vectors, and denote
it by $T_1^{()}=\left\{0,1\right\}^{{\ell}}$. This set is
partitioned to $m_1$ equally sized sub-codes
$T_2^{(0)},T_2^{(1)},...,T_2^{(m_1-1)}$, and  each sub-code
$T_2^{(b_1)}$ is in turn partitioned to $m_2$ equally sized codes
$T_3^{(b_1,0)},T_3^{(b_1,1)},...,T_3^{(b_1,m_2-1)}$
($b_1\in\left\{0,1,...,m_1-1\right\}$). This partitioning may be
further carried on.

\begin{definition} \label{def: compoIntoCosets}
The set $\left\{T_1,...,T_m\right\}$ is called a decomposition of $\left\{0,1\right\}^{\ell}$ , if
 $T_1^{()} = \left\{0,1 \right\}^{\ell}$, and $T_i^{({\bf b}_1^{i-1})}$ is partitioned into $m_i$ equally sized sets $\left\{T_{i+1}^{({\bf b}_1^{i-1}b_i)}\right\}_{b_i =0,1,...,m_i-1}$, of size $\frac{2^{{\ell}}}{\prod_{j=1}^{i}m_j}$ ($i\in[m-1]$). We denote the set of sub-codes of level number $i$ by $$T_i = \left\{ T_i^{({\bf b}_1^{i-1})}|b_j\in \left\{0,1,2,...,m_j-1\right\},j\in [i-1]\right\}.$$ The partition is usually  described by the following chain of codes parameters $$(n_1,k_1,d_1)-(n_2,k_2,d_2)-...-(n_m,k_m,d_m),$$ if for each $\mathcal{T}\in T_i$ we have that  $\mathcal{T}$ is a code of length $n_i$, size $2^{k_i}$ and minimum distance at least $d_i$.

 If the sub-codes of the decompositions are cosets, then we say that $\left\{T_1,...,T_m\right\}$ is a decomposition into cosets. In this case, for each $T_i$ the sub-code that contains the zero codeword is called the representative sub-code, and a minimal weight codeword for each coset  is called the coset leader. If all the sub-codes in the decomposition are cosets of linear codes, we say that the decomposition is linear.
\end{definition}
\begin{example}
As an example consider ${\ell}=4$ and the $4\times 4$ binary matrix
$$
G = \left(
      \begin{array}{cccc}
        1 & 0 & 0 & 0 \\
        1 & 1 & 0 & 0 \\
        1 & 0 & 1 & 0 \\
        1 & 1 & 1 & 1 \\
      \end{array}
    \right).
$$
A partition into cosets, having the following chain of parameters
$(4,4,1)-(4,3,2)-(4,1,4)$,  can be implied by the matrix. This is
done by taking $T_1^{()} = \left\{0,1\right\}^4$, which is
partitioned to the even weight codewords and odd weight codewords
cosets, i.e. $T_2^{(0)}=\left\{{\bf u}_1^4| \sum_{i=1}^{4}u_i \equiv
0 (\mbox{\rm mod} \, 2) \right\}$, $T_2^{(1)}=\left\{{\bf u}_1^4|
\sum_{i=1}^{4}u_i \equiv 1 (\mbox{\rm mod} \, 2) \right\}$, these
cosets are in turn partitioned to anti podalic pairs,
$T_3^{(0,0)}=\{0000,1111\}$, $T_3^{(0,1)}=\{1010,0101\}$,
$T_3^{(0,2)}=\{1100,0011\}$, $T_3^{(0,3)}=\{0110,1001\}$, and
$T_3^{(1,b)}=[1 0 0 0 ]+T_3^{(0,b)}$ ($b\in\left\{0,1,2,3\right\}$).
Note, that in order to describe this partition, it suffices to
describe the representatives and the coset leaders for the partition
of the representatives.
\end{example}
A binary transformation can be associated to a code decomposition in the following way.
\begin{definition}\label{def: transAssiciatedTheCodes}
Let $\left\{T_1,T_2,...,T_{{\ell}+1}\right\}$ be a code decomposition of $\left\{0,1\right\}^{\ell}$, such that $m_i=2$ for each $i\in[{\ell}]$. Note that the code $T_i^{\left({\bf b}_1^{i-1}\right)}$ is of size $2^{{\ell}-i+1}$, and specifically  $T_{{\ell}+1}^{{(b_1,b_2,...,b_{{\ell}})}}$ contains only one codeword. We call such a decomposition a binary decomposition.
The  transformation $g\left(\cdot\right):\left\{0,1\right\}^{\ell}\rightarrow\left\{0,1\right\}^{\ell}$ induced by this binary code decomposition is defined as follows.
\begin{equation}\label{eq:gBinDefined}
g({\bf u}_1^{\ell})={\bf x}_1^{\ell} \,\,\,\,\, \text{if  } {\bf
x}_1^{\ell} \in T_{{\ell}+1}^{\left({\bf u}_1^{{\ell}}\right)}.
\end{equation}
\end{definition}

Following the definition, we can observe, that a sequential decision
making on the bits of the input to the transformation (${\bf
u}_1^{{\ell}}$) given a noisy observation of the output is actually
a decision on the sub-code to which the transmitted vector belongs
to. As such, deciding on the first bit $u_1$ is actually deciding if
the transmitted vector belongs to $T_2^{(0)}$ or to $T_2^{(1)}$.
Once we decided on $u_1$, we assume that we transmitted a codeword
of $T_2^{(u_1)}$ and by deciding on $u_2$ we choose the appropriate
refinement or sub-code of $T_2^{(u_1)}$, i.e. we should decide
between the candidates $T_3^{(u_1,0)}$ and $T_3^{(u_1,1)}$. Due to
this fact, it comes as no surprise that  the Hamming distances
between two candidate sub-codes plays an important role when
considering the rate of polarization.
\begin{definition}\label{def:partialDistances}
For a binary code decomposition as in Definition \ref{def:
transAssiciatedTheCodes}, the Hamming distances between sub-codes in
the decomposition are defined as follows:

$$D_{min}^{(i)}({\bf u}_1^{i-1})=
\min\left\{d_H({\bf c}_1,{\bf c}_2)\Big| {\bf c}_1\in
 T_{i+1}^{\left({\bf u}_1^{i-1}\cdot 0\right)},{\bf c}_2\in T_{i+1}^{\left({\bf u}_1^{i-1}\cdot
 1\right)}\right\},$$
$$
D_{min}^{(i)} = \min\left\{D_{min}^{(i)}({\bf u}_1^{i-1})\big|{\bf
u}_1^{i-1}\in \left\{0,1\right\}^{i-1} \right\}.
$$
\normalsize
\end{definition}

A transformation $g\left(\cdot\right)$ can be used as a building block for a recursive construction of a transformation of greater length, in a similar manner to \cite{Arikan}. We specify this construction explicitly in the next definition.
\begin{definition}\label{def:constructG}
Given a transformation $g(\cdot)$ of dimension ${\ell}$, we
construct a mapping $g^{(m)}(\cdot)$ of dimension ${\ell}^m$ (i.e.
$g^{(m)}(\cdot):\left\{0,1\right\}^{{\ell}^m}\rightarrow\left\{0,1\right\}^{{\ell}^m}$)
in the following recursive fashion.
$$g^{(1)}({\bf u}_1^{\ell})=g({\bf u}_1^{\ell})\,\,\,;$$

$$g^{(m)}=\Big[ g^{(m-1)}\left(\gamma_{1,1}, \gamma_{2,1}, \gamma_{3,1}, \ldots, \gamma_{{\ell}^{m-1},1}\right),$$
$$\,\,\,\,\,\,\,g^{(m-1)}\left(\gamma_{1,2}, \gamma_{2,2}, \gamma_{3,2}, \ldots, \gamma_{{\ell}^{m-1},2}\right),\ldots,$$
$$
\,\,\,\,\,\,\,g^{(m-1)}\left(\gamma_{1, {\ell}}, \gamma_{2, {\ell}}, \gamma_{3, {\ell}}, \ldots, \gamma_{{\ell}^{m-1},{\ell}}\right)  \Big],
$$
\normalsize
where
$$
\gamma_{i,j}=g_j\left({\bf u}_{(i-1)\cdot {\ell} +1}^{i\cdot {\ell}}\right)
\,\,\,\,\,  1 \leq i \leq {\ell}^{m-1} \,\,\,\,\,\, 1 \leq j \leq
{\ell}.
$$
\end{definition}
The transformation $g^{(m)}(\cdot)$ can be used to transmit data
over the B-MC channel. The method of successive cancelation can now
be used to decode, with decoding complexity of $O\left(2^{\ell}\cdot
N\cdot \log_{\ell}(N)\right)$ as in \cite{Arikan}.

We use the same channel definition, the corresponding symmetric
capacity and the Bhattacharyya parameter as in
\cite{Arikan,Korada,MoriandTanka}. Note that for uniform binary
random vectors $U_1^{\ell}$, and
$X_1^{\ell}=g\left(U_1^{\ell}\right)$ we have that
$I(Y_1^{\ell};U_1^{\ell})=I(Y_1^{\ell};X_1^{\ell})$, because the
transformation $g(\cdot)$ is invertible. Furthermore, since we consider
memoryless channels, we have $I(Y_1^{\ell};X_1^{\ell}) = {\ell}\cdot
I(Y_1;X_1) = {\ell}\cdot I(W)$, and on the other hand
$$
I(Y_1^{\ell};U_1^{\ell})  = \sum_{i=1}^{\ell}
I(Y_1^{\ell};U_i|U_{1}^{i-1}) = \sum_{i=1}^{\ell} I(W^{(i)}).
$$
Define the tree process of the channels generated by the kernels, in
the same way as it was done in \cite{Arikan} and generalized in
\cite{Korada}. A random sequence $\left\{W_n\right\}_{n\geq 0}$ is
defined such that $W_{n} \in \left\{W^{(i)} \right\}_{i=1}^{{\ell}^n}$
with
$$
W_0 = W
$$
$$
W_{n+1}=W_n^{(B_{n+1})},
$$
where $\left\{B_n\right\}_{n\geq 1}$ is a sequence of i.i.d random variables uniformly distributed over the set $\left\{0,1,2,...,{\ell}-1\right\}$. In a similar manner, the symmetric capacity corresponding to the channels  $\left\{I_n\right\}_{n\geq 0} =\left\{I(W_n)\right\}_{n\geq 0} $ and the Bhattacharyya parameters random variables  $\left\{Z_n\right\}_{n\geq 0 } =\left\{Z(W_n)\right\}_{n\geq 0 }$ are defined. Just as in \cite[Proposition 8]{Arikan}, we can prove that the random sequence $\left\{I_n\right\}_{n\geq 0}$ is a bounded martingale, and it is uniform integrable which means it converges almost surely to $I_{\infty}$ and that $\mathbb{E}\left\{I_{\infty} \right\} = I(W)$. Now, if we can show that $Z_n \rightarrow Z_{\infty}$ w.h.p such that $Z_{\infty} \in \left\{0,1 \right\}$, by the relations between the channel's information and the Bhattacharyya parameter \cite[Proposition 1]{Arikan}, we have that $I_{\infty} \in \left\{0,1\right\}$. But, this means that $\Pr\left(I_{\infty}  = 1\right)=\mathbb{E}\left\{ I_{\infty} \right\} = I(W)$, which is the channel polarization phenomenon.
\begin{proposition}\label{propo: polarizationOfg}
Let $g(\cdot)$ be a binary transformation of dimension ${\ell}$,
induced by a binary code decomposition
$\left\{T_1,T_2,...,T_{{\ell}+1}\right\}$. If  there exists ${\bf
u}_1^{{\ell}-1}\in\left\{0,1\right\}^{{\ell}-1}$  such that
$D_{min}^{(\ell)}({\bf u}_1^{{\ell}-1})\geq 2$, then
$\Pr\left(I_{\infty} = 1\right)=I(W)$.
\end{proposition}
\proof In \cite[Corollary 11]{MoriandTanka},
sufficient conditions are given for
\begin{equation}\label{eq:prop1E1}
\lim_{n\rightarrow \infty}\Pr\left(Z_n \in \left(\delta,1-\delta \right) \right) =0\,\,\,\,\,\forall \delta\in (0,0.5).
\end{equation}
The first condition is that there exists a vector ${\bf u}_1^{{\ell}-1}$, indices
$i,j\in [{\ell}]$ and permutations $\sigma(\cdot)$, and
$\tau(\cdot)$ on $\left\{0,1\right\}$ such that
$$
g^{({\bf u}_1^{{\ell}-1})}_{i}({  u}_{\ell})=\sigma({
u}_{\ell})\,\,\,\,\,\,\,\text{and}\,\,\,\,\,\,\, g^{({\bf
u}_1^{{\ell}-1})}_{j}({  u}_{\ell})=\mu({  u}_{\ell}).
$$
This requirement applies here, because if there exists ${\bf
u}_1^{{\ell}-1}\in\left\{0,1\right\}^{{\ell}-1}$  such that
$D_{min}^{(\ell)}({\bf u}_1^{{\ell}-1})\geq 2$, then the two codewords
of the code $T_{\ell}^{({\bf u}_1^{{\ell}-1})}$, ${\bf c}_1$ and
${\bf c}_2$, are at  Hamming distance at least 2. This means that
there exist  at least two indices $i,j$ such that $c_{1,i}\neq
c_{2,i}$ and $c_{1,j}\neq c_{2,j}$, therefore $g^{({\bf
u}_1^{{\ell}-1})}_{i}({  u}_{\ell})$ and $g^{({\bf
u}_1^{{\ell}-1})}_{j}({   u}_{\ell})$ are both permutations. The
second condition is that for any ${\bf v}_1^{{\ell}-1}\in\left\{0,1
\right\}^{{\ell}-1}$ there exist an index $m\in[{\ell}]$ and a permutation
$\mu(\cdot)$ on $\left\{0,1\right\}$ such that
$$
g^{({\bf v}_1^{{\ell}-1})}_{m}({  v}_{\ell})=\mu({ v}_{\ell}).
$$
This requirement also applies here, by noting that for each ${\bf
v}_1^{{\ell}-1}\in\left\{ 0,1\right\}^{{\ell}-1}$ the two codewords
of the set $T_{\ell}^{({\bf v}_1^{{\ell}-1})}$ are at  Hamming
distance at least 1. This means that (\ref{eq:prop1E1}) holds, which
implies that $I_{\infty} \in \left\{0,1\right\}$ almost surely, and
therefore $\Pr\left(I_{\infty}  = 1\right)=I(W)$. \ \qed

The next proposition on the rate of polarization is an easy consequence of \cite[Theorem 19]{MoriandTanka} and Proposition \ref{propo: polarizationOfg}.
\begin{proposition}\label{propo:rateOfPolarization}
Let $g(\cdot)$ be a bijective transformation of dimension ${\ell}$, induced by
code partitioning $\left\{T_1,T_2,...,T_{{\ell}+1}\right\}$. If
there exists ${\bf u}_1^{{\ell}-1}\in\left\{0,1\right\}^{{\ell}-1}$
such that $D_{min}^{(\ell)}({\bf u}_1^{{\ell}-1})\geq 2$, then

(i) For any $\beta<E(g)$
$$
    \lim_{n\rightarrow \infty} \Pr\left(Z_n\leq 2^{-{\ell}^{n\beta}}\right)=I(W),
$$

(ii) For any $\beta>E(g)$
    $$
    \lim_{n\rightarrow \infty} \Pr\left(Z_n\geq 2^{-{\ell}^{n\beta}}\right)=1,
    $$
    where $E(g) = \frac{1}{{\ell}}\sum_{i=1}^{{\ell}}\log_{\ell}\left(D_{min}^{(i)}\right)$.
\end{proposition}

Naturally, we would like to find kernels maximizing $E(g)$. In the next section we consider upper bounds on the maximum achievable exponent per dimension $\ell$.
\section{Bounds on the Optimal Exponent}\label{sec:Bounds}
We define the optimal exponent per dimension $\ell$ as
\begin{equation}\label{eq:El}
E_{\ell}=\max_{g:\{0,1\}^{\ell}\rightarrow\{0,1\}^{\ell}}E\left(g\right).
\end{equation}
Note that in \cite{Korada}, $E_{\ell}$ was defined as a maximization over the set of binary linear kernels, and here we extend the definition for general kernels. Furthermore, a lower bound on the kernel using Gilbert-Vershamov technique also applies in this case \cite[Lemma 20]{Korada}. The following lemma is a generalization of \cite[Lemma 18]{Korada}.
\begin{lemma}[Generalization of \cite{Korada}, Lemma 20]\label{lem:increasingD}
Let $g:\{0,1\}^{\ell} \rightarrow \{0,1\}^{\ell}$ be a polarizing kernel. Fix $k\in[\ell-1]$ and define a mapping \begin{equation}
\tilde{g}\left({\bf v}_1^{\ell}\right)=g\left({\bf v}_1^{k-1} \,\,\,,v_{k+1} \,,\,v_k\,\,,\, {\bf v}_{k+2}^{\ell}\right),
\end{equation}
i.e in this mapping the coordinates $k$ and $k+1$ are swapped. Let $\left\{D_{\text{min}}^{(i)}\right\}_{i=1}^{\ell}$ and $\left\{\tilde{D}_{\text{min}}^{(i)}\right\}_{i=1}^{\ell}$ denote the partial distances of $g(\cdot)$ and $\tilde{g}(\cdot)$ respectively. If $D_{\text{min}}^{(k)}>D_{\text{min}}^{(k+1)}$ then
\begin{enumerate}[label=(\roman{*})]
  \item $E(g)\leq E(\tilde{g})$
  \item $\tilde{D}_{\text{min}}^{(k)}<\tilde{D}_{\text{min}}^{(k+1)}$
\end{enumerate}
\end{lemma}
\proof
We follow the path of the proof of \cite[Lemma 20]{Korada}.
It will be useful to introduce the following equivalent definition of the partial distance sequence.
\begin{equation}\label{eq:partialDistRedef}
D_{\text{min}}^{(i)}=\min\Big\{d_H\left(g\left({\bf w}_{1}^{i-1},0,{\bf u}_{i+1}^{\ell}\right),g\left({\bf w}_{1}^{i-1},1,{\bf v}_{i+1}^{\ell}\right)\right)\Big|{\bf w}_{1}^{i-1},{\bf u}_{i+1}^{\ell},{\bf v}_{i+1}^{\ell} \Big\}
\end{equation}
According to this definition it is easy to see that
\begin{equation}
D_{\text{min}}^{(i)}={\tilde D}_{\text{min}}^{(i)}\,\,\,\,\, i\in [\ell]\backslash\{k,k+1\}.
\end{equation}
Hence, it suffices to show that
\begin{equation}\label{eq:SuffCond}
D_{\text{min}}^{(k)}\cdot D_{\text{min}}^{(k+1)}\leq {\tilde{D}}_{\text{min}}^{(k)}\cdot {\tilde{D}}_{\text{min}}^{(k+1)}
\end{equation}
in order to prove (i).

Using (\ref{eq:partialDistRedef}), we have
\begin{equation}\label{eq:Dk}
  D_{\text{min}}^{(k)}=\min\Big\{d_H\left(g\left({\bf w}_{1}^{k-1},0,{\bf u}_{k+1}^{\ell}\right),g\left({\bf w}_{1}^{k-1},1,{\bf v}_{k+1}^{\ell}\right)\right)\Big|{\bf w}_{1}^{k-1},{\bf u}_{k+1}^{\ell},{\bf v}_{k+1}^{\ell} \Big\}
  \end{equation}
  \begin{equation}\label{eq:tDk}
    {\tilde{D}}_{\text{min}}^{(k)}=\min\Big\{d_H\left(g\left({\bf w}_{1}^{k-1},u_k,0,{\bf u}_{k+2}^{\ell}\right),g\left({\bf w}_{1}^{k-1},v_k,1,{\bf v}_{k+2}^{\ell}\right)\right)\Big|{\bf w}_{1}^{k-1},{\bf u}_{k+2}^{\ell},{\bf v}_{k+2}^{\ell},u_k,v_k \Big\}
    \end{equation}
    \begin{equation}\label{eq:Dkp1}
  D_{\text{min}}^{(k+1)}=\min\Big\{d_H\left(g\left({\bf w}_{1}^{k},0,{\bf u}_{k+2}^{\ell}\right),g\left({\bf w}_{1}^{k},1,{\bf v}_{k+2}^{\ell}\right)\right)\Big|{\bf w}_{1}^{k},{\bf u}_{k+2}^{\ell},{\bf v}_{k+2}^{\ell} \Big\}.
  \end{equation}
  \begin{equation}\label{eq:tDkp1}
  {\tilde{D}}_{\text{min}}^{(k+1)}=\min\Big\{d_H\left(g\left({\bf w}_{1}^{k-1},0,w_{k+1},{\bf u}_{k+2}^{\ell}\right),g\left({\bf w}_{1}^{k-1},1,w_{k+1},{\bf v}_{k+2}^{\ell}\right)\right)\Big|{\bf w}_{1}^{k-1},w_{k+1},{\bf u}_{k+2}^{\ell},{\bf v}_{k+2}^{\ell} \Big\}.
\end{equation}
Because the set on which we perform the minimization in (\ref{eq:tDkp1}) is a subset of the set on which we preform the minimization in  (\ref{eq:Dk}) we have that $  D_{\text{min}}^{(k)}\leq {\tilde{D}}_{\text{min}}^{(k+1)} $. On the other hand, the minimization in (\ref{eq:tDk}) can be expressed as $ {\tilde{D}}_{\text{min}}^{(k)}=\min\Big\{\Delta_1,\Delta_2\Big\}$, where
\begin{equation}\label{eq:tDk2a}
\Delta_1 =  \min\Big\{d_H\left(g\left({\bf w}_{1}^{k-1},w_k,0,{\bf u}_{k+2}^{\ell}\right),g\left({\bf w}_{1}^{k-1},w_k,1,{\bf v}_{k+2}^{\ell}\right)\right)\Big|{\bf w}_{1}^{k-1},{\bf u}_{k+2}^{\ell},{\bf v}_{k+2}^{\ell},w_k \Big\}
\end{equation}

\begin{equation}\label{eq:tDk2b}
\Delta_2 =  \min\Big\{d_H\left(g\left({\bf w}_{1}^{k-1},w_k,0,{\bf u}_{k+2}^{\ell}\right),g\left({\bf w}_{1}^{k-1},1-w_k,1,{\bf v}_{k+2}^{\ell}\right)\right)\Big|{\bf w}_{1}^{k-1},{\bf u}_{k+2}^{\ell},{\bf v}_{k+2}^{\ell},w_k \Big\}.
\end{equation}
We see that $\Delta_1=  D_{\text{min}}^{(k+1)}$ and $\Delta_2\geq D_{\text{min}}^{(k)}$. So,  ${\tilde{D}}_{\text{min}}^{(k)}=D_{\text{min}}^{(k+1)}$, because $D_{\text{min}}^{(k)}>D_{\text{min}}^{(k+1)}$. So this proves (\ref{eq:SuffCond}) and therefore (i).
Now,
$$
{\tilde{D}}_{\text{min}}^{(k)}=D_{\text{min}}^{(k+1)}<D_{\text{min}}^{(k)}\leq {\tilde{D}}_{\text{min}}^{(k+1)},
$$
which results in (ii). \qed

Lemma \ref{lem:increasingD} implies that when seeking the optimal exponent,$E_{\ell}$, for a given dimension $\ell$, it suffices to consider kernels with non-decreasing partial distance sequences.
This observation also results in \cite[Lemma 22]{Korada}
\begin{lemma}[\cite{Korada},Lemma 22]\label{lem:upperBound}
Let $d(n,k)$ denote the largest possible minimum distance of a binary code of length $n$ and size $2^k$. Then,
\begin{equation}\label{eq:UBKorada1}
E_{\ell}\leq \frac{1}{\ell}\sum_{i=1}^{\ell}\log_{\ell}\left(d(\ell,\ell-i+1)\right)
\end{equation}
\end{lemma}
\proof
Consider a polarizing kernel $g(\cdot)$ having partial distance sequence $\left\{D_{\text{min}}^{(i)}\right\}_{i=1}^{\ell}$. Because of Lemma \ref{lem:increasingD}, we can assume that the sequence is non decreasing (otherwise, we can find a kernel that is having a non-decreasing sequence with at least the same exponent). Note that
\begin{equation}\label{eq:UBKorada1prf}
D_{\text{min}}^{(k)} = \min_{i\geq k} D_{\text{min}}^{(i)}=\min_{{\bf u}_1^{k-1}}\Big\{\min\big\{d_H({\bf c}_1, {\bf c}_2)\Big | {\bf c}_1,{\bf c}_2\in T_{k}^{\left({\bf u}_1^{k-1}\right)},{\bf c}_1\neq {\bf c}_2 \big\}\Big\}\leq d(\ell,\ell-k+1),
\end{equation}
where the second inequality is due to the fact that each of the codes in the inner minimum, (i.e. $T_{k}^{\left({\bf u}_1^{k-1}\right)}$), is of size $2^{\ell-k+1}$ and length $\ell$. \qed

As already noted in \cite{Korada}, the shortcoming of (\ref{eq:UBKorada1}) as an upper-bound, is that the dependencies between the partial distances are not exploited. For binary and linear kernels, \cite[Lemma 26]{Korada} gives an improved upper bound utilizing these dependencies. In the sequel we develop an upper bound  that is applicable to general kernels. The basic idea of the bound we develop, is to express the partial distance sequence of a kernel, in terms of distance distributions of a code.

For a code $\mathcal{C}$ of length $\ell$ and size $M$ we define the distance distribution as
\begin{equation}
B_i=\frac{1}{M}\left|\left\{({\bf c}_1,{\bf c}_2)\big| d_H({\bf c}_1,{\bf c}_2)=i \right\} \right| \,\,\,\,\ 0\leq i \leq \ell.
\end{equation}
Note that $B_0 = 1$ and
\begin{equation}\label{eq:Req0}
\sum_{i=1}^{\ell}B_i=M-1.
\end{equation}
Now, given a non decreasing partial distance sequence $\left\{D_{\text{min}}^{(i)}\right\}_{i=1}^{\ell}$ we choose an arbitrary $k\in [\ell]$ and consider the sub-sequence $\left\{D_{\text{min}}^{(i)}\right\}_{i=k}^{\ell}$. Using the reasoning that led to (\ref{eq:UBKorada1prf}), we  observe that we need to consider the sub-codes $\left\{T_{k}^{\left({\bf u}_1^{k-1}\right)}\right\}_{{\bf u}_1^{k-1}\in\{0,1\}^{k-1}}$ of size $M=2^{\ell-k+1}$, but whereas in (\ref{eq:UBKorada1}) we considered only the minimum distance, here we  may have additional requirements from the distance distribution of the code. Let's begin by understanding the meaning of $D_{\text{min}}^{(\ell)}$ (the last element of the sequence). By definition, the code $T_{k}^{\left({\bf u}_1^{k-1}\right)}$ is decomposed into $\frac{2^{\ell-k+1}}{2}$ sub-codes of size $2$, such that in each one the distance between the $2$ codewords is at least $D_{\text{min}}^{(\ell)}$. This means that we must fulfill the following requirement
\begin{equation}\label{eq:Req1}
\sum_{i=D_{\text{min}}^{(\ell)}}^{\ell} B_i\geq 1,
\end{equation}
where $\left\{B_o\right\}_{i=0}^{\ell}$ is the distance distribution  of  $T_{k}^{\left({\bf u}_1^{k-1}\right)}$.
Now, let's proceed to $D_{\text{min}}^{(\ell-1)}$. This item implies that there are $\frac{2^{\ell-k+1}}{2^2}$ sub-codes of $T_{k}^{\left({\bf u}_1^{k-1}\right)}$ of $4$ codewords that each one of them can be decomposed into $2$ sub-codes of $2$ code-words having minimum distance between the sub-codes of at least $D_{\text{min}}^{(\ell-1)}$. From this, we deduce that there are  $2\cdot 2^{\ell-k+1}$ pairs of codewords having their distance at least $D_{\text{min}}^{(\ell-1)}$. These pairs are an addition to the the ones we counted in (\ref{eq:Req1}). Thus, because we assume that the partial distance sequence is non-decreasing, we have the following requirement.
\begin{equation}\label{eq:Req2}
\sum_{i=D_{\text{min}}^{(\ell-1)}}^{\ell} B_i\geq 3.
\end{equation}
Note that if $D_{\text{min}}^{(\ell-1)}=D_{\text{min}}^{(\ell)}$ then (\ref{eq:Req1}) is redundant given (\ref{eq:Req2}).
In the general case, when considering $D_{\text{min}}^{(\ell-r)}$, where $0 \leq r \leq \ell-k$, we take into account
$\frac{2^{\ell-k+1}}{2^{r+1}}$ sub-codes of $T_{k}^{\left({\bf u}_1^{k-1}\right)}$, each one of size $2^{r+1}$ and each one can be partitioned into two sub-codes of which the minimum distance between them is $D_{\text{min}}^{(\ell-r)}$. So, there are
$
2\cdot \frac{2^{\ell-k+1}}{2^{r+1}}\cdot \left(2^{r}\right)^2 = M \cdot 2^r
$
 codewords pairs (that were not counted at the previous steps)  such that their distance is at least $D_{\text{min}}^{(\ell-r)}$. Summarizing, we get the following set of $\ell-k+1$ inequalities
\begin{equation}\label{eq:ReqGen}
 \sum_{i=D_{\text{min}}^{(\ell-r)}}^{\ell} B_i\geq \sum_{j=0}^{r}2^j = 2^{r+1}-1  \,\,\,\,\,\,\,\,\,\,\,\,\, 0\leq r\leq \ell-k.
\end{equation}

By Delsarte \cite{Delsarte73}, we can specify additional linear requirements on the distance distribution, by
\begin{equation}\label{eq:ReqDelsarte}
\sum_{j=1}^{\ell}B_j\cdot P_i(j) \geq -{\ell  \choose i} \,\,\,\,\,\,\,\,\, 0\leq i \leq \ell,
\end{equation}
where $P_k(x)$ is the Krawtchouk polynomial, which is defined as
\begin{equation}\label{eq:Kraw}
P_k(x)=\sum_{m=0}^{k}(-1)^m{x \choose m}{\ell - x \choose k-m}.
\end{equation}
In addition, the following is also an obvious requirement
\begin{equation}\label{eq:ReqNonNegative}
B_i \geq 0 \,\,\,\,\,\,\, i\in [\ell].
\end{equation}
We see that  requirements (\ref{eq:Req0}),(\ref{eq:ReqGen}),(\ref{eq:ReqDelsarte}) and (\ref{eq:ReqNonNegative}) are all linear. A partial distance sequence that corresponds to a kernel must be able to  fulfill these requirements for every $k\in [\ell]$. So, taking the maximum exponent corresponding to a partial distance sequence that fulfils the requirement for each $k\in[\ell]$ results in an upper-bound on the exponent.  Checking the validity of a sequence can be done by linear programming methods (we need to check if the polytope is not empty). We now turn to give two simple examples of the method, and after them we present a variation on this development that leads to a stronger bound.
\begin{example}
Consider $\ell = 3$. Let $\left\{D_{\text{min}}^{(i)}\right\}_{i=1}^{3}$ be the partial distance sequence. Note first that by the singelton bound $D_{\text{min}}^{(k)}\leq k$. We first consider the possibility that $D_{\text{min}}^{(3)}=3$ and $D_{\text{min}}^{(2)}=2$. This assumption is translated by (\ref{eq:Req0}) and (\ref{eq:ReqGen}) to
\begin{equation}\label{eq:exl3_1}
B_2+B_3=3\,\,\,\,\,\,\,\,\,\,, B_3\geq 1 \,\,\,\,\,,\,\,\,\, B_2,B_3\geq 0
\end{equation}
By (\ref{eq:ReqDelsarte}) for $i=1$ we have
$$
B_2\cdot P_1(2)+B_3\cdot P_1(3) \geq -3
$$

\begin{equation}\label{eq:exl3_2}
-B_2-3\cdot B_3 \geq -3\,\,\,\, \Longrightarrow_{\text{(\ref{eq:exl3_1})}}  B_2 =0,B_3 = 3
\end{equation}
on the other hand for (\ref{eq:ReqDelsarte}) $i=3$ we have
$$
B_2-B_3 \geq -1
$$
which is a contradiction to (\ref{eq:exl3_2}).
The next best candidate is a sequence having $D_{\text{min}}^{(2)}=D_{\text{min}}^{(3)}=2$, this sequence is feasible by considering the following generating matrix
$$
\left(
  \begin{array}{ccc}
    1 & 0 & 0 \\
    1 & 1 & 0 \\
    0 & 1 & 1 \\
  \end{array}
\right).
$$
This proves that $E_{3}=\frac{1}{3}\log_3{4}\approx0.42062$.
\end{example}

\begin{example}
Consider $\ell = 4$. Let $\left\{D_{\text{min}}^{(i)}\right\}_{i=1}^{4}$ be the partial distance sequence. We first consider the possibility that $D_{\text{min}}^{(4)}=D_{\text{min}}^{(3)}=3$  (if this possibility is eliminated it means that $D_{\text{min}}^{(4)}=4,D_{\text{min}}^{(3)}=3$ is also not possible).
  (\ref{eq:Req0}) and (\ref{eq:ReqGen}) are translated to
\begin{equation}\label{eq:exl4_1}
B_3+B_4=3\,\,\,\,\, \,\,\,\, B_3,B_4\geq 0
\end{equation}
By (\ref{eq:ReqDelsarte}) for $i=1$ we have
$$
B_3\cdot P_1(3)+B_4\cdot P_1(4) \geq -4
$$
$$
-2\cdot B_3-4\cdot B_4 \geq -4\Longrightarrow_{\text{(\ref{eq:exl4_1})}}B_3 +2(3-B_3)\leq 2\Longrightarrow B_3\geq 4
$$
which is a contradiction to (\ref{eq:exl4_1}).
The next best candidate is $$D_{\text{min}}^{(4)}=4,D_{\text{min}}^{(3)}=2,D_{\text{min}}^{(2)}=2,D_{\text{min}}^{(1)}=1,$$
which can be achieved by a binary linear kernel induced by the generating matrix
$$
\left(
  \begin{array}{cc}
    1 & 0 \\
    1 & 1 \\
  \end{array}
\right)^{\otimes2}.
$$
This proves that $E_4=0.5$.
\end{example}

The idea of transforming the partial distance sequence into requirements on distance distributions can be further refined.
As we did before, we begin our discussion  by considering the sub-sequence $\left\{D_{\text{min}}^{(i)}\right\}_{i=k}^{\ell}$. We start by giving an interpretation to $D_{\text{min}}^{(\ell)}$ (the last element of the sequence). By definition, the code $T_{k}^{\left({\bf u}_1^{k-1}\right)}$ is decomposed into $\frac{2^{\ell-k+1}}{2}$ sub-codes of size $2$, where in each one the distance between the $2$ codewords is at least $D_{\text{min}}^{(\ell)}$. Denote by $B_{i}^{\left({\bf u}_1^{\ell-1}\right)} \,\,\,\, i\in [\ell]$ the partial distance distribution of the sub-code $T_{k}^{\left({\bf u}_1^{\ell-1}\right)}$ of the code $T_{k}^{\left({\bf u}_1^{k-1}\right)}$. By definition we have
\begin{equation}
B_{i}^{\left({\bf u}_1^{\ell-1}\right)} = \frac{1}{2}\left|\left\{d_H({\bf c}_1,{\bf c}_2 )=i\Big| {\bf c}_1,{\bf c}_2\in T_{\ell}^{\left({\bf u}_1^{\ell-1}\right)}\right\}\right|.
\end{equation}
Obviously,
\begin{equation}
\sum_{i=D_{\text{min}}^{(\ell)}}^{\ell} B_{i}^{\left({\bf u}_1^{\ell-1}\right)} = 1 \,\,\,\,\, \forall {\bf u}_k^{\ell-1}\in\{0,1\}^{\ell-k},
\end{equation}
\begin{equation}
\sum_{j=1}^{\ell}B_{j}^{\left({\bf u}_1^{\ell-1}\right)}\cdot P_i(j) \geq -{\ell  \choose i} \,\,\,\,\,\,\,\,\, 0\leq i \leq \ell,\forall {\bf u}_k^{\ell-1}\in\{0,1\}^{\ell-k}.
\end{equation}
Define the average of this distribution over all the sub-codes of $T_{k}^{\left({\bf u}_1^{k-1}\right)}$, i.e.
\begin{equation}
\bar{B}_{i}^{(\ell)}=\frac{1}{2^{\ell-k}}\sum_{{\bf u}_k^{\ell-1}\in\{0,1\}^{\ell-k}}B_{i}^{\left({\bf u}_1^{\ell-1}\right)} \,\,\,\, i\in[\ell].
\end{equation}
Note that
\begin{equation}
\bar{B}_{i}^{(\ell)} = \frac{1}{M}\left|\left\{d_H({\bf c}_1,{\bf c}_2 )=i\Big| {\bf c}_1,{\bf c}_2\in T_{k}^{\left({\bf u}_1^{\ell-1}\right)}, {\bf u}_k^{\ell-1}\in\{0,1\}^{\ell-k}\right\}\right|
\end{equation}
and
\begin{equation}
\sum_{i=D_{\text{min}}^{(\ell)}}^{\ell}\bar{B}_{i}^{(\ell)} = 1 ,
\end{equation}
\begin{equation}
\sum_{j=1}^{\ell}\bar{B}_{j}^{(\ell)}\cdot P_i(j) \geq -{\ell  \choose i} \,\,\,\,\,\,\,\,\, 0\leq i \leq \ell.
\end{equation}
Let's proceed to $D_{\text{min}}^{(\ell-1)}$.
By definition, the code $T_{k}^{\left({\bf u}_1^{k-1}\right)}$ is decomposed into $\frac{2^{\ell-k+1}}{4}$ sub-codes of size $4$, where in each one the distance between the $2$ codewords is at least $D_{\text{min}}^{(\ell-1)}$. Denote by $B_{i}^{\left({\bf u}_1^{\ell-2}\right)} \,\,\,\, i\in [\ell]$, the  distance distribution of the sub-code $T_{k}^{\left({\bf u}_1^{\ell-2}\right)}$ of the code $T_{k}^{\left({\bf u}_1^{k-1}\right)}$.
\begin{equation}
B_{i}^{\left({\bf u}_1^{\ell-2}\right)} = \frac{1}{4}\left|\left\{d_H({\bf c}_1,{\bf c}_2 )=i\Big| {\bf c}_1,{\bf c}_2\in T_{\ell-1}^{\left({\bf u}_1^{\ell-2}\right)}\right\}\right|.
\end{equation}
Note that
\begin{equation}
B_{i}^{\left({\bf u}_1^{\ell-2}\right)}\geq \frac{1}{2}\left(B_{i}^{\left({\bf u}_1^{\ell-2}\cdot0\right)}+B_{i}^{\left({\bf u}_1^{\ell-2}\cdot1\right)}\right).
\end{equation}
So by introducing the average distance distribution
\begin{equation}
\bar{B}_{i}^{(\ell-1)}=\frac{1}{2^{\ell-k-1}}\sum_{{\bf u}_k^{\ell-2}\in\{0,1\}^{\ell-k-1}}B_{i}^{\left({\bf u}_1^{\ell-2}\right)} \,\,\,\,\,\,\,\, i\in[\ell],
\end{equation}
we get
\begin{equation}
\sum_{i=D_{\text{min}}^{(\ell-1)}}^{\ell}\bar{B}_{i}^{(\ell-1)} = 3 ,
\end{equation}
\begin{equation}
\sum_{j=1}^{\ell}\bar{B}_{j}^{(\ell-1)}\cdot P_i(j) \geq -{\ell  \choose i} \,\,\,\,\,\,\,\,\, 0\leq i \leq \ell.
\end{equation}
and
\begin{equation}
\bar{B}_{i}^{(\ell-1)} - \bar{B}_{i}^{(\ell)}\geq 0 \,\,\,\,\,\,\,\,\, 0\leq i \leq \ell.
\end{equation}
In the general case, when taking $D_{\text{min}}^{(\ell-r)}$ into account, where $0 \leq r \leq \ell-k$, we essentially consider the $\frac{2^{\ell-k+1}}{2^{r+1}}$ sub-codes of $T_{k}^{\left({\bf u}_1^{k-1}\right)}$, each one of size $2^{r+1}$ and each one can be partitioned into two sub-codes of size $2^r$ of which the minimum distance between them is $D_{\text{min}}^{(\ell-r)}$. Denote the distance distribution of the sub-code $T_{\ell -r}^{\left({\bf u}_1^{\ell -(r+1)}\right)}$ as  $\left\{B_{i}^{\left({\bf u}_1^{\ell-(r+1)}\right)}\right\}_{i\in[\ell]}$ and the average distance distribution  as $\left\{\bar{B}_{i}^{(\ell-r)}\right\}_{i\in[\ell]}$. We have
\begin{equation}
B_{i}^{\left({\bf u}_1^{\ell-(r+1)}\right)} = \frac{1}{2^{r+1}}\left|\left\{d_H({\bf c}_1,{\bf c}_2 )=i\Big| {\bf c}_1,{\bf c}_2\in T_{\ell-r}^{\left({\bf u}_1^{\ell-(r+1)}\right)}\right\}\right|,
\end{equation}
\begin{equation}
\bar{B}_{i}^{(\ell-r)}=\frac{1}{2^{\ell-k-r}}\sum_{{\bf u}_k^{\ell-(r-1)}\in\{0,1\}^{\ell-k-r}}B_{i}^{\left({\bf u}_1^{\ell-(r+1)}\right)} \,\,\,\,\,\,\,\,\,, i\in[\ell],
\end{equation}

which results in
\begin{equation}
 \sum_{i=D_{\text{min}}^{(\ell-r)}}^{\ell}\bar{B}_{i}^{(\ell-r)}= \sum_{j=0}^{r}2^j = 2^{r+1}-1
\end{equation}

\begin{equation}
\bar{B}_{i}^{(\ell-r)}-\bar{B}_{i}^{(\ell-r+1)}\geq 0
\end{equation}
\begin{equation}
\sum_{j=1}^{\ell}\bar{B}_{j}^{(\ell-r)}\cdot P_i(j) \geq -{\ell  \choose i} \,\,\,\,\,\,\,\,\, 0\leq i \leq \ell.
\end{equation}

We summarize this development.

\begin{definition}
Let $\left\{ D_i \right\}_{i\in[\ell]}$ be a monotone non-increasing sequence of non-negative integral numbers, such that $ D_i \leq d(\ell,\ell-i+1)$.
We say that this sequence is $\ell$ dimension \textit{Linear Programming (LP) valid} if the polytope defined by the following non negative variables $\left\{\bar{B}_i^{(k)}\big| 1\leq k\leq \ell ,\,\,\,\,\, D_{k} \leq i \leq \ell \right\} $ is not empty.
\begin{equation}
 \sum_{i=D_{\ell-r}}^{\ell}\bar{B}_{i}^{(\ell-r)}= \sum_{i=0}^{r}2^i = 2^{r+1}-1\,\,\,\,\,\, 0\leq r\leq \ell-1
\end{equation}

\begin{equation}
\bar{B}_{i}^{(\ell-r)}-\bar{B}_{i}^{(\ell-r+1)}\geq 0\,\,\,\,\,\,\, 1\leq r \leq \ell-1,\,\,\,\,\,D_{\ell-r+1} \leq i \leq \ell
\end{equation}
\begin{equation}
\sum_{j=D_{\ell-r}}^{\ell}\bar{B}_{j}^{(\ell-r)}\cdot P_i(j) \geq -{\ell  \choose i} \,\,\,\,\,\,\,\,\, 0\leq i \leq \ell,\,\,\,\,\,\,\,\,0\leq r\leq \ell-1
\end{equation}
\end{definition}

\begin{proposition}\label{propo:LPvalid}
If  $\left\{ D_{\text{min}}^{(i)}\right\}_{i\in[\ell]}$ is a partial distance sequence corresponding to some binary $\ell$ dimension kernel $g(\cdot)$, then $\left\{ D_{\text{min}}^{(i)}\right\}_{i\in[\ell]}$  is $\ell$-dimension LP-valid sequence.
\end{proposition}

We denote by $\mathcal{V}^{(\ell)}_{\text{LP}}$ the set of $\ell$-dimension $LP$-valid sequences. The following proposition is an easy consequence of Proposition \ref{propo:LPvalid}.
\begin{proposition}\label{propo:UBLP}
\begin{equation}\label{eq:UBOur}
E_{\ell}\leq \max_{\left\{D_k \right\}_{k\in[\ell]} \in \mathcal{V}^{(\ell)}_{\text{LP}}} \frac{1}{\ell}\sum_{i=1}^{\ell} \log_{\ell}D_i.
\end{equation}
\end{proposition}
It should be noted that the method of Proposition \ref{propo:UBLP} can be easily generalized to non-binary kernels using the appropriate (non-binary) Krawtchouk polynomials.
We computed the bound for several instances of $\ell$ by carefully enumerating the sequences in $\mathcal{V}^{(\ell)}_{\text{LP}}$ using Wolfram's \textit{Mathematica} LP-Solver. Table \ref{tbl:UBEl} contains the results for $12\leq\ell\leq 16$.
In the
next section, we give examples of good kernels, that are derived by utilizing results about known code decompositions, for $14\leq\ell\leq16$ that achieve the optimal exponent.
\begin{table}
\center
\begin{tabular}{|c|c|l|c|}
  \hline
  $\#$ &${\ell}$&optimal sequence & $E_{\ell}$  \\
  \hline
  1 & 12  & $1, 2, 2, 2, 2, 4, 4, 4, 6, 6, 6, 12$  & $0.49605$ \\
  2 & 13 & $1, 2, 2, 2, 2, 4, 4, 4, 6, 6, 6, 8, 10$ & $0.500498$ \\
  3 & 14 & $1, 2, 2, 2, 2, 4, 4, 4, 6, 6, 6, 8, 8, 8$ & 0.50194 \\
  4 & 15 & $1, 2, 2, 2, 2, 4, 4, 4, 6, 6, 6, 8, 8, 8, 8$ & $0.507733$ \\
  5 & 16 & $1, 2, 2, 2, 2, 4, 4, 4, 6, 6, 6, 8, 8, 8, 8, 16$ & $0.52742$ \\
  \hline
\end{tabular}
 \normalsize
\caption{$E_{\ell}$ per different dimensions }\label{tbl:UBEl}
\end{table}

\section{Designing Kernels by  Known Code Decompositions}\label{sec:ExmpOfGoodKernels}
As we noticed in  Section \ref{sec:CodesDecom}, the  exponent, $E(g)$, is influenced by Hamming distances between the subsets in the binary partition $\left\{T_1,...,T_{\ell+1}\right\}$. In this section, we use a particular method for getting good distances by using known decompositions, which are not necessarily binary decompositions. The following observation links between general decompositions and  binary decompositions.

\begin{observation}\label{obs:linkBetSetPartCosetDecomo}
If there exists a code decomposition of $\left\{0,1\right\}^{\ell}$ with the following chain of parameters
$$
({\ell},k_1,d_1)-({\ell},k_2,d_2)-...-({\ell},k_m,d_m),
$$
then there exists a binary code decomposition of $\left\{0,1\right\}^{\ell}$, such that
$$
D_{min}^{(i)} \geq d_{j}\,\,\,\,\, \text{where} \,\,\,\, k_{j+1}<
{\ell}-i+1 \leq k_{j},\,\,\,\,\,\,\,
$$
$$
j\in[m],\,\,i\in[{\ell}],\,\,
k_{m+1}=0.
$$
\end{observation}

The next observation about the kernel exponent is an easy consequence of the previous observation.
\begin{observation}\label{obs:linkBetSetPartCosetAndErrorExponent}
If there exists a code decomposition of $\left\{0,1\right\}^{\ell}$
with the following chain of parameters
$$
({\ell},k_1,d_1)-({\ell},k_2,d_2)-...-({\ell},k_m,d_m),
$$
then there exists an ${\ell}$ dimensional binary kernel $g(\cdot)$
induced by a binary code decomposition
$\left\{T_1,...,T_{{\ell}+1}\right\}$ such that
\begin{equation}\label{eq:partCosetsE1LB}
E(g) \geq (1/{\ell})\cdot \sum_{i=1}^{m} (k_i-k_{i+1})\cdot \log_{\ell}\left(d_i\right),
\end{equation}
where $k_{m+1}=0$.
\end{observation}

In \cite[Table 5]{LitsynTblBinaryCodes}, the author gives a list of code  decompositions for ${\ell}\leq 16$.
Using this list, we can construct polarizing non-linear kernels and get lower bounds on their  exponent $E(g)$ (In order to do so, we use  Observation \ref{obs:linkBetSetPartCosetAndErrorExponent} and Propositions \ref{propo: polarizationOfg} and \ref{propo:rateOfPolarization}). Table \ref{tbl:errorExp} contains a list of code decompositions that give lower bounds on $E(g)$ that are greater than 0.5. At the chain
description column of the table, the code length equals ${\ell}$ for all the sub-codes,
and was omitted from the chain for brevity. Note that the second entry of the table has the exponent of the kernel suggested in \cite{Korada}. It was proven that this is the best linear binary kernel of dimension 16, and that all the linear kernels of dimension $<16$ have exponents $\leq 0.5$. The first entry of the table gives a non-linear decomposition resulting in a non linear kernel having a better exponent. In fact, this exponent is even better than  all the exponents that were recorded in \cite[Table 1]{Korada}. Furthermore, entries $1,3$ and $4$ achieve the optimal exponent per their dimension as Table \ref{tbl:UBEl} indicates. Thus, the exponent value indicated in Table \ref{tbl:errorExp} is not just a lower bound, but rather the true exponent.
The appendix contains details about the decompositions in Table \ref{tbl:errorExp}.
\begin{table}
\center
\scriptsize
\begin{tabular}{|c|c|l|c|}
  \hline
  $\#$ &${\ell}$&chain description&lower  \\
     & & &  bound on\\
     & & &  $E(g)$\\
  \hline
  1 & 16  & $(16,1)-(15,2)-(11,4)-(8,6)-(5,8)-(1,16)$  & 0.52742 \\
  2 & 16 & $(16,1)-(15,2)-(11,4)-(7,6)-(5,8)-(1,16)$ & 0.51828 \\
  3 & 15 & $(15,1)-(14,2)-(10,4)-(7,6)-(4,8)$ & 0.50773 \\
  4 & 14 & $(14,1)-(13,2)-(9,4)-(6,6)-(3,8)$ & 0.50193 \\
  \hline
\end{tabular}
 \normalsize
\caption{Code decompositions from \cite[Table
5]{LitsynTblBinaryCodes} with their corresponding lower bounds on
kernel exponents for the kernels induced by them. }\label{tbl:errorExp}
\end{table}

\section{Conclusions}
The notion of code decomposition was used for the design of good binary kernels in the sense of the polar code exponent. Some of the kernels we suggested are proven to achieve the optimal exponent per their dimension. It should be noted that by using non-binary kernels one can get better exponents, as was demonstrated in \cite{MoriandTanka3}. There is an essential loss, when using non-binary code decomposition for designing binary kernels. It seems that if we allow the inputs of the kernel to be from different alphabet sizes, we may gain an additional improvement. This interesting idea is further explored in a sequel paper by the authors \cite{PrShLi3}.
\ifloguseIEEEConf
    \appendix
\else
    \section*{Appendix}
\fi
In this appendix we give  details on the decompositions enumerated in Table \ref{tbl:errorExp}. All of the decompositions are coset decompositions, so we only need to specify the sub-code representatives.

\subsubsection*{\#1)$(16,16,1)-(16,15,2)-(16,11,4)-(16,8,6)-(16,5,8)-(16,1,16)$}
The sub-code representatives are  $(16,15,2)$  single parity check code,  $(16,11,4)$ extended Hamming code, $(16,8,6)$ Nordstrom-Robinson code, $(16,5,8)$ first order Reed-Muller code, $(16,1,16)$ repetition code.
\subsubsection*{\#2)$(16,16,1)-(16,15,2)-(16,11,4)-(16,7,6)-(16,5,8)-(16,1,16)$}
The sub-code representatives are  $(16,15,2)$ - single parity check
code,  $(16,11,4)$ - extended Hamming code, $(16,7,6)$ -   extended
$2$-error correcting BCH code, $(16,5,8)$-   first-order Reed-Muller
code, $(16,1,16)$ - repetition code.
\subsubsection*{\#3)$(15,15,1)-(15,14,2)-(15,10,4)-(15,7,6)-(15,4,8)$}
The sub-code representatives are $(15,14,2)$ - single parity check
code,  $(15,10,4)$ - shortened extended Hamming code, $(15,7,6)$ - shortened
Nordstrom-Robinson code, $(15,4,8)$ - shortened first order
Reed-Muller code.
\subsubsection*{\#4)$(14,14,1)-(14,13,2)-(14,9,4)-(14,6,6)-(14,3,8)$}
The sub-code representatives are $(14,13,2)$ - single parity check
code,  $(14,9,4)$ - twice shortened extended Hamming code,
$(14,6,6)$ - twice shortened Nordstrom-Robinson code, $(14,3,8)$ -
twice shortened first order Reed-Muller code.

\subsubsection*{Explicit Encoding of Decomposition $\#1$}
For decomposition $\#1$ we elaborate on the kernel mapping function $g(\cdot):\left\{0,1\right\}^{16}\rightarrow\left\{0,1\right\}^{16}$. To do so, we use Table \ref{tbl:elabCodeDecomp}.
The third column from the left determines whether the vectors on the second column are all the coset vectors (if they do not form a linear space) or just the basis for the space of coset vectors (if they form a linear space). The fourth and the fifth columns determine the stage of the code decomposition these vectors belong to;  the "main code" is decomposed to cosets of the "sub-code" (each coset is generated by adding a different coset vector from the set specified by column 2 to the sub-code). The entry corresponding to indices $9-11$ is taken from \cite{LitsynFastDecoding}.

We now describe the encoding process. Let ${\bf u}_{1}^{16}$ be a binary vector. The indices of the vector are partitioned to subsets according to the first column of the table.  For each subset the corresponding sub-vector of $\bf u$ is mapped to a coset vector. The mapping can be arbitrary, but when the coset vectors form a linear space, we usually prefer to multiply the corresponding sub-vector by a generating matrix which rows are the vectors in the "coset vectors" column. To get the value of $g({\bf u})$, we add-up the six coset vectors we got from the last step. Note that using this mapping definition, it is easy to derive the mapping function corresponding to decompositions $\#3$ and $\#4$ as well.

\begin{table}
\scriptsize
\center
\begin{tabular}{| c|c|c|c|c|}
  \hline
  input   & coset vectors &  coset vectors  & main code & sub-code
  \\
             vector  &              &    form a &      &
               \\
                 indices     &              &         linear space?  &   &
               \\
  \hline
  $1$ & $[0	0	0	0	0	0	0	0	0	0	0	0	0	 0	0	 1
]$ & yes & $(16,16,1)$ & $(16,15,2)$ \\
\hline
  $2-5$ & $[0	0	0	0	0	0	0	1	0	0	0	0	 0	0	0	 1
]$ & yes & $(16,15,2)$ & $(16,11,4)$ \\
    & $[0	0	0	0	0	0	0	0	0	0	0	1	0	 0	0	 1
]$ &   &   &   \\
    & $[0	0	0	0	0	0	0	0	0	0	0	0	0	 1	0	 1
]$ &   &   &   \\
    &  $[0	0	0	0	0	0	0	0	0	0	0	0	0	 0	1	 1
]$ &   &   &   \\
\hline
  $6-8$ & $[0	0	0	1	0	0	0	1	0	0	0	1	 0	0	0	 1
]$ & yes & $(16,11,4)$ & $(16,8,6)$ \\
   & $[0	0	0	0	0	1	0	1	0	0	0	0	0	 1	0	 1]$
 &   &   &   \\
   & $[0	0	0	0	0	0	0	0	0	1	0	1	0	 1	0	 1
]$ &   &   &   \\
\hline
  $9-11$ & $[0	0	0	0	0	0	0	0	0	0	0	0	 0	0	0	 0
] $& no &  $(16,8,6)$ & $(16,5,8)$ \\
  & $[0	0	0	0	0	0	1	1	0	1	0	1	0	1	 1	0
]$ &   &   &   \\

    & $[0	0	0	1	0	0	0	1	0	1	0	0	1	 0	1	 1
]$ &   &   &   \\
       & $[0	0	0	1	0	0	1	0	0	0	1	0	 1	1	1	 0
]$ &   &   &   \\
   & $[0	0	0	1	0	1	1	1	0	0	0	1	1	 0	0	 0
]$ &   &   &   \\
  & $[0	0	0	0	0	1	1	0	0	0	1	1	0	1	 0	1
]$ &   &   &   \\
    & $[0	0	0	1	0	1	0	0	0	1	1	1	0	 0	1	 0
]$ &   &   &   \\
  & $[0	0	0	0	0	1	0	1	0	1	1	0	1	1	 0	0
]$ &   &   &   \\
\hline
  $13-15$ & $[0	1	0	1	0	1	0	1	0	1	0	1	 0	1	0	 1
]$ & yes & $(16,5,8)$ & $(16,1,16)$ \\
    & $[0	0	1	1	0	0	1	1	0	0	1	1	0	 0	1	 1
]$ &  &  &  \\
   & $[0	0	0	0	1	1	1	1	0	0	0	0	1	 1	1	 1
]$ &   &   &   \\
    & $[0	0	0	0	0	0	0	0	1	1	1	1	1	 1	1	 1
]$ &   &   &   \\
\hline
  $16$ & $[1	1	1	1	1	1	1	1	1	1	1	1	 1	1	1	 1]$ &  yes &  $(16,1,16)$ & -  \\
  \hline
\end{tabular}
\normalsize
\caption{Coset vectors for code decomposition $\#1$.   }\label{tbl:elabCodeDecomp}
\end{table}

\bibliographystyle{IEEEtran}
\bibliography{IEEEabrv,bibTexPolar}

%
%
%

\end{document}